# Unipolar Vertical Transport in GaN/AlGaN/GaN Heterostructures

D. N. Nath, P.S. Park, Z. C. Yang & S. Rajan

The Ohio State University, Department of Electrical and Computer Engineering
Columbus, OH, 43210

**Abstract:**

In this letter, we report on unipolar vertical transport characteristics in c-plane GaN/AlGaN/GaN heterostructures. Vertical current in heterostructures with random alloy barriers was found to be independent of dislocation density and heterostructure barrier height, and significantly higher than theoretical estimates. Percolation-based transport due to random alloy fluctuations in the ternary AlGaN is suggested as the dominant transport mechanism, and confirmed through experiments showing that non-random or digital AlGaN alloys and polarization-engineered binary GaN barriers can eliminate percolation transport and reduce leakage significantly. The understanding of vertical transport and methods for effective control proposed here will greatly impact III-nitride unipolar vertical devices.

Recently, there has been renewed interest in unipolar III-nitride vertical devices such as resonant tunneling diodes (RTD)[1,2,3] hot electron transistors (HET)[4,5], tunnel-injection hot electron transfer amplifiers (THETA)[6,7,8] . III-nitride vertical transistors, in particular, are of interest since heterojunction bipolar transistors, which are very successful in the InP/GaAs material system, are not viable due to the low hole mobility and lifetime in GaN. However, high vertical leakage currents in n-GaN/AlGaN/n-GaN heterostructures have held back the development of unipolar vertical III-nitride transistors. In this letter, we propose that percolation based leakage due to alloy fluctuations in the ternary AlGaN barrier dominates vertical transport, and show that non-random barriers can provide control of vertical transport. The results presented here are expected to have important impact on a range of III-nitride devices such as HEMTs and LEDs, as well as vertical transport devices such as current aperture vertical electron transfer (CAVET)[9,10] transistors, resonant tunneling devices (RTD), and hot electron transistors.

To study unipolar transport characteristic in a III-nitride heterostructure, we investigate an epitaxial stack as shown in Figure 1(a) which consists of a 15-20 nm of degenerately doped n+-GaN top contact layer and an $Al_xGa_{1-x}N$ barrier 30-35 nm thick. The n+-GaN template on which samples are grown serves as the bottom n-contact. The epitaxial stacks for the devices were grown by plasma-assisted molecular beam epitaxy (MBE) using a Veeco 930 system equipped with a uni-bulb Veeco $N_2$ plasma source. Commercially available Ga-polar free-standing n-doped GaN substrates (TDD ~ $5x10^7$ $cm^{-2}$) from St. Gobain were used for the growths. High-resolution X-ray diffractometer scans were done to verify the thicknesses and compositions of epitaxial layers. Al (20 nm)/Ni (20 m)/Au (100 nm)/Ni (20 nm) were e-beam evaporated for top-contact metal stack. An inductively coupled plasma/re-active ion (ICP-RIE) etch recipe using 40 V RIE and 40 W ICP power with 50 sccm $Cl_2$/5 sccm $BCl_3$ was used to achieve a mesa isolation depth of 100 nm with a controlled etch rate. Thereafter bottom contact metal stack Al (20 nm)/Ni (20 m)/Au (100 nm)/Ni (20 nm) was evaporated. Depending on the device, mesa area was between 90 to 120 μm$^2$. Devices were probed using sub-micron probe tips and measured using an Agilent B1500 Semiconductor Parameter Analyzer.

To investigate the effect of heterostructure barrier height on reverse bias leakage, a series of three samples with Al-composition 16%, 27% and 37% in the AlGaN barrier were grown, with thickness fixed at 30 nm. As seen in the energy band diagrams (Figure 1b) obtained using self-consistent 1-D Schrodinger-Poisson simulations[11], AlGaN barrier with 16% Al-composition provides a heterojunction barrier in excess of 0.5 eV from the base Fermi level at equilibrium. The reverse bias leakage current due to Fowler-Nordheim tunneling associated with such a heterostructure with 16% Al-composition was estimated theoretically using analytical calculations, and using Atlas Silvaco modeling[12]. As seen in Figure 1(c), theoretically, negligible leakage is expected up to reverse bias of 3 V. However the typical current densities measured experimentally on devices fabricated were found to be orders of magnitude higher. Figure 1d shows the measured vertical I-V characteristics of the three samples with 16%, 27% and 37% Al-compositions in the barrier. The magnitude of leakage current in reverse bias was similar for these three samples despite a significant difference in their heterostructure barrier heights. While both thermionic and tunnel current would theoretically be expected to vary by orders of magnitude with such changes in the barrier height, our experiments show that the AlGaN composition had negligible effect on the vertical current leakage in GaN/AlGaN/GaN heterostructures. We can therefore discount tunneling or thermionic emission mechanisms. Temperature dependence of the IV characteristics (Figure 2) revealed that vertical conductivity did not vary significantly with temperature. We conclude that the origin of leakage in heterostructures is not a thermally activated mechanism such as trap-assisted tunneling[13] or thermionic emission over the barrier.

Threading dislocation density (TDD), particularly related to screw type dislocations, is widely believed to impact[14,15] reverse bias leakage in AlGaN/GaN Schottky diodes and HEMTs[16]. However, the impact of dislocation density on semiconductor heterojunction leakage has not been investigated. To investigate the effect of dislocations, epitaxial structures were grown on a low-dislocation density bulk GaN substrate[17] (Sample LD, TDD < $10^5$ cm$^{-2}$) and a higher dislocation density HVPE grown substrate[18] (Sample HD, TDD ~ $5 \times 10^7$ cm$^{-2}$). Both samples were grown and processed together to eliminate growth and process variations. Figure 3a and 3b show typical 5 μm x 5 μm Atomic Force Microscope (AFM) scans of surface morphologies of the epitaxial stacks for Samples HD and LD respectively. In Sample HD, ~30 dislocations (black pits), usually associated with screw type dislocations, were observed in the scan area, corresponding to TDD ~ $10^8$ cm$^{-2}$. No pits associated with dislocations were visible in the AFM micrograph for Sample LD, indicating TDD < $10^5$ cm$^{-2}$. I-V characteristics for GaN/AlGaN/GaN structures (Figure 3c) revealed that despite three orders of magnitude difference in TDDs, the vertical currents observed in Samples LD and HD were very similar. This shows that TDDs are not the primary factor responsible for the anomalous high currents in unipolar GaN/AlGaN/GaN heterostructures.

Since neither heterojunction barrier height nor a reduction in TDDs affect the behavior of GaN/AlGaN/GaN heterostructures, and since there is almost no temperature dependence in the current density, our observations suggest that electrons see *almost zero effective barrier* to transport across the heterojunction. We propose that percolation-based transport through the Ga-rich regions of AlGaN is the dominant current mechanism in all the samples described earlier (Figure 4a). Such percolation-based transport can be attributed to composition fluctuations in ternary alloy layer[18] due to statistical distribution of Ga and Al atoms in group-III sites. To directly test the percolation-based transport hypothesis, two separate barrier configurations were explored that both eliminate random alloy fluctuations, and were compared with random $Al_{0.3}Ga_{0.7}N$ barriers (Sample R). The first approach used was a digital AlGaN barrier (Sample D) with alloy barrier grown by repetition of 2 ML AlN/ 4 ML GaN digital periods. Such an approach (Figure 4c) eliminates the possibility of statistical fluctuations that could lead to Ga-rich regions in the AlGaN. The number of digital periods was such that the barrier corresponded to 25-30 nm of AlGaN with average composition 25% verified by dynamic XRD simulations. Figures 4b shows the energy band diagrams corresponding to Sample D showing that the effective barrier height (based on the minibands in the superlattice) in this case was similar to that for ternary AlGaN.

We designed a second non-random alloy barrier based on a polarization-engineered binary GaN barrier (sample B1) (Figure 4 (d)). As shown in the energy band diagram (Figure 4(e)), a thin InGaN[19] layer provides a polarization-induced dipole that creates an *electrostatic* barrier that would not be permeable to percolation effects.

Figures 5a and 5b show the vertical I-V characteristics of the random alloy barrier (R), the digital alloy barrier (D1) and the polarization engineered binary GaN barrier (B1). The samples with non-random digital (D1) and binary GaN (B1) barriers had 2-to-3 orders of magnitude lower leakage current density than the sample R (random alloy) for reverse bias in the range of < 2 V, even though the effective barrier height from the energy band diagram in these cases is nominally the same. The low reverse bias leakage characteristic is not top-contact limited since the lateral current between two top-contact pads of a device exhibits linear Ohmic behavior (inset to Figure 5a). This significant reduction in vertical leakage using a binary GaN and a digital alloy (Al$_{.3}$Ga$_{.7}$N) as barriers confirms our hypothesis that eliminating a ternary random alloy as the barrier *does* prevent percolation-based GaN-to-GaN nano-island hopping transport of electrons.

The results presented here are of great significance to vertical unipolar III-nitride devices since excess leakage in such structures had prevented them from achieving their theoretical performance. In addition, our work is also significant in understanding forward and reverse bias gate characteristics for III-Nitride AlGaN/GaN HEMTs and metal-insulator transistor HEMTs, and electron barrier layers in III-nitride LEDs. In the case of MISHEMTs, forward bias on the gate has been found to create a large electron accumulation at the insulator/oxide interface even at low effective forward voltage where tunneling through the AlGaN should be minimal. Our work on understanding and eliminating percolation transport through the barrier could help to eliminate these effects and enable normally off power transistors that operate at positive gate bias.

In conclusion, we investigated unipolar transport in GaN/AlGaN/GaN heterostructures and found the reverse bias leakage to be independent of TDDs as well as of heterojunction potential heights. It is hypothesized that random alloy fluctuations in the ternary AlGaN barrier lead to percolation-based transport which enables electrons to flow through the ternary barrier. This hypothesis is verified and vertical leakage is reduced by 2-to-3 orders of magnitude by using non-random alloy based barriers based on digital Al$_{.3}$Ga$_{.7}$N and polarization-engineered binary GaN barriers. This understanding of unipolar transport characteristics and means of reducing the vertical leakage in GaN/AlGaN/GaN heterostructures a variety of existing as well as novel unipolar III-nitride devices, as well as for other devices such as HEMTs and LEDs that use random AlGaN alloy-based electron blocking layers.

**Acknowledgment**: This work is funded by the Office of Naval Research (Dr. Paul Maki) [DATE MURI].

**Figure Legends**:

**Figure 1: Effect of heterojunction barrier height on leakage**

(a) Typical epitaxial stack of devices used in this study (b) Energy band diagrams of device structures with $Al_xGa_{1-x}N$ barrier with x = 0.16, 0.27 and 0.37 (c) Comparison of calculated, simulated and experimentally measured vertical leakage current density corresponding to an $Al_{.16}Ga_{.84}N$ barrier (d) Comparison of measured leakage for samples with $Al_xGa_{1-x}N$ barrier with x = 0.16, 0.27 and 0.37

**Figure 2:** Temperature-dependent vertical leakage current showing weak temperature dependence

**Figure 3: Effect of dislocation density on leakage**

5 µm x 5 µm AFM scan showing surface morphology of epitaxial stack of device grown on

(a) St. Gobain substrates. The number of black pits correspond to TDD ~ $10^8$ cm$^{-2}$. Data scale = 3 nm. (b) Ammono substrate. Data scale = 2.5 nm. (c) Comparison of measured leakage for devices on St. Gobain and Ammono substrates, showing that a difference of 3-orders of magnitude in TDDs had no role in the leakage.

**Figure 4** (a) Schematic showing GaN-to-GaN nano-island mediated percolative transport in a ternary random alloy (AlGaN) layer (b) Energy band diagram sample D1 with a digital AlGaN barrier (c) Schematic showing digital AlGaN which can prevent percolative transport (d) Epitaxial stack for the sample where a binary GaN barrier is explored (e) Energy band diagram for the same.

**Figure 5:** Comparison of measured leakage for samples with random alloy, binary GaN and digital alloys as barriers in (a) linear (b) log scale. Inset to Figure 5(a) shows the lateral I-V between two top-contacts of the same device indicating its linear and Ohmic nature (the digital alloy sample).

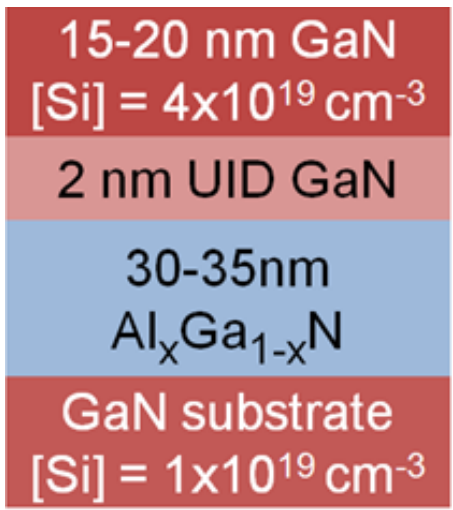

Figure 1(a)

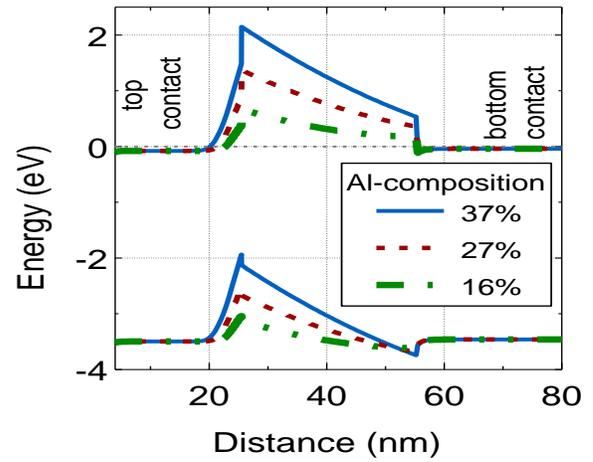

Figure 1(b)

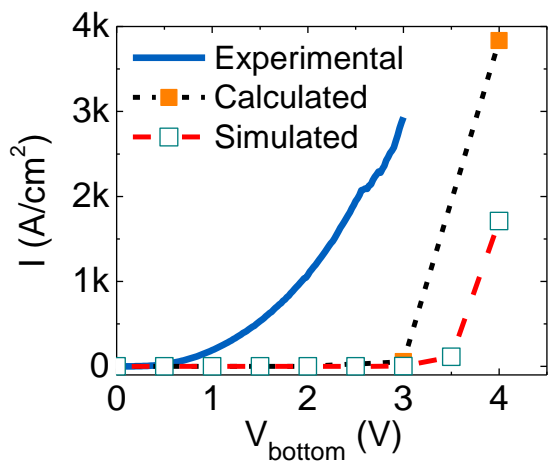

Figure 1(c)

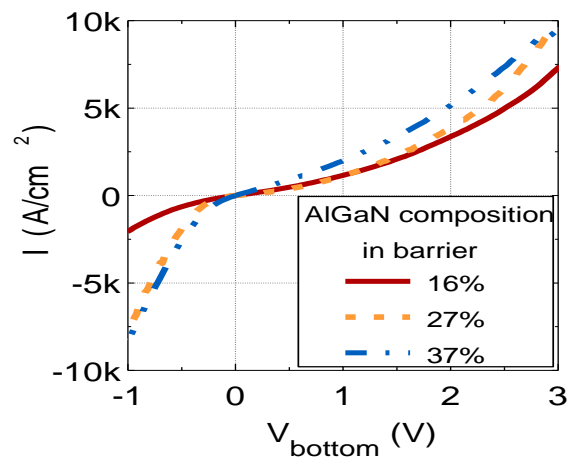

Figure 1(d)

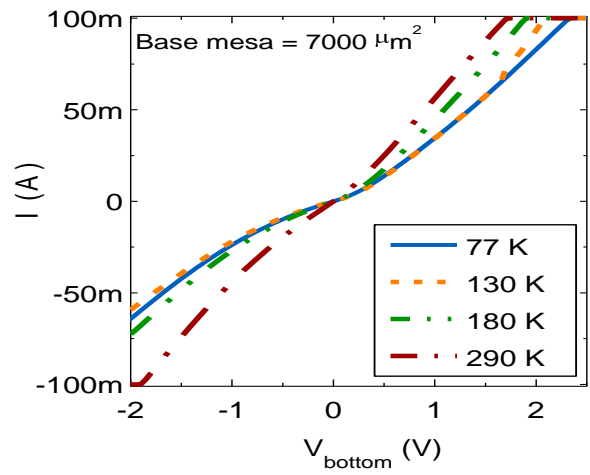

**Figure 2**

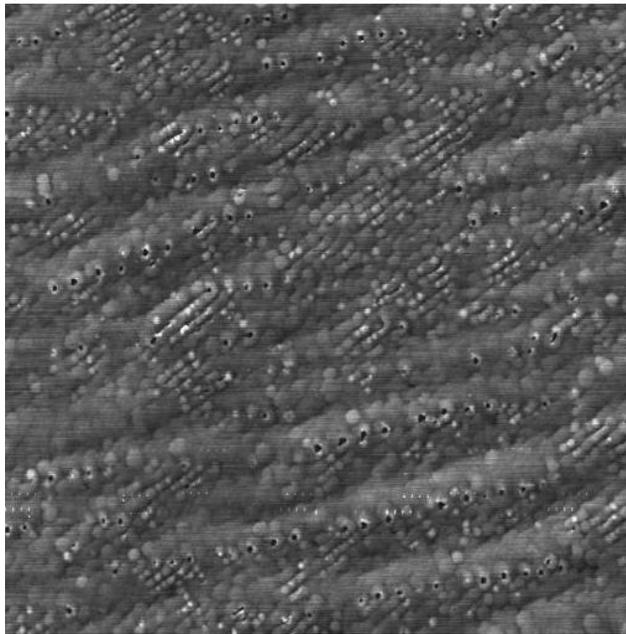
**Figure 3(a)**

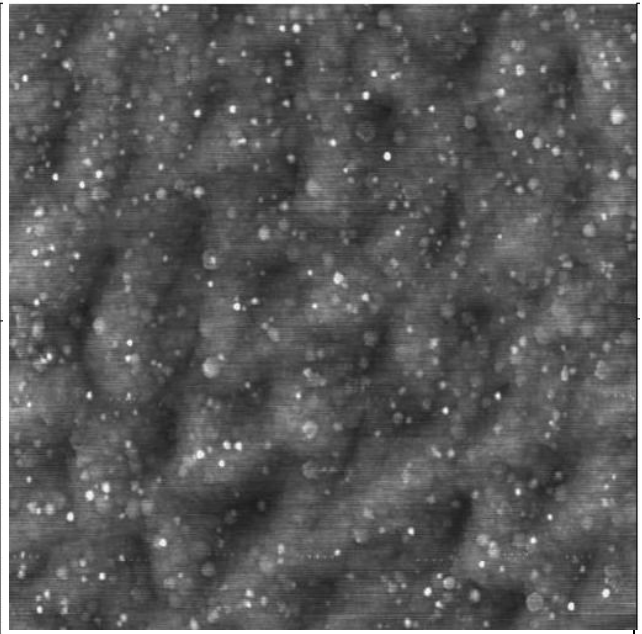
**Figure 3(b)**

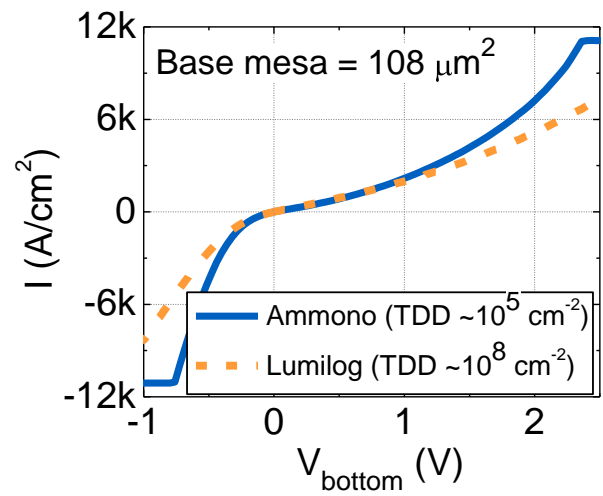

**Figure 3(c)**

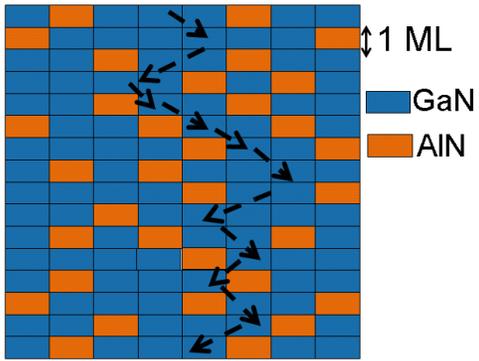

**Figure 4(a)**

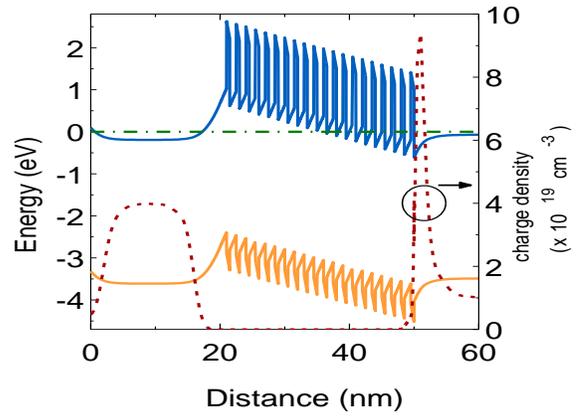

**Figure 4(b)**

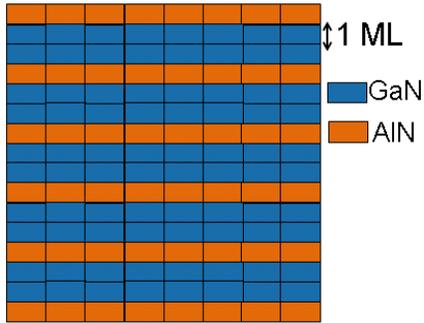

**Figure 4(c)**

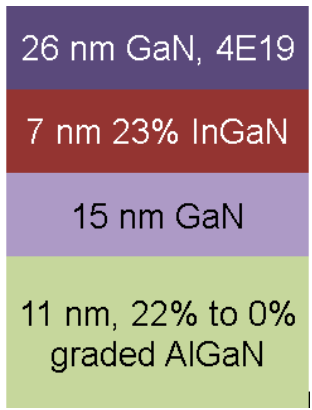

**Figure 4(d)**

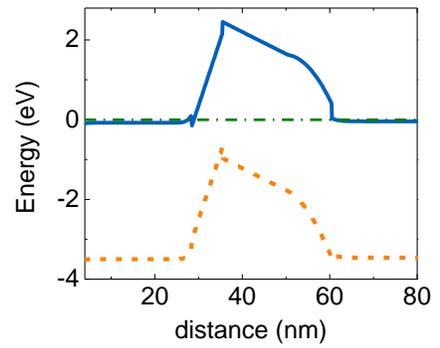

**Figure 4(e)**

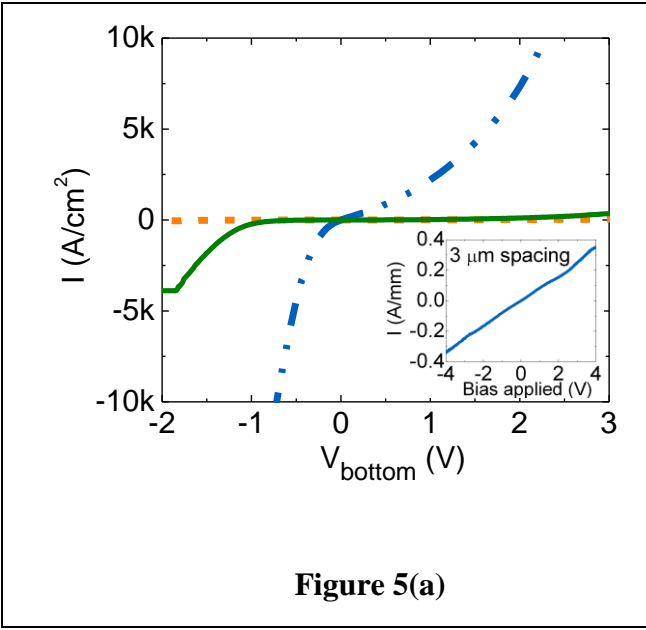 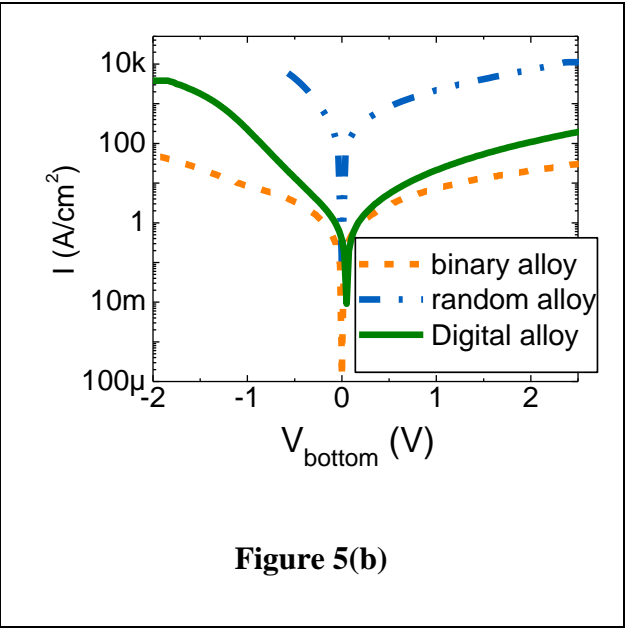

**Figure 5(a)** **Figure 5(b)**